\documentclass[preprint,12pt,fleqn]{elsarticle}
\usepackage{amssymb}
\usepackage{amsfonts}
\usepackage[english]{babel}
\usepackage{euscript}
\usepackage{bbm}
\usepackage{xfrac}

\usepackage{color}
\usepackage{mathrsfs}
\usepackage{calrsfs}
\usepackage{amsbsy}

\journal{Journal of Geometry and Physics}

\newcommand{\fl}{\hspace*{-\mathindent}}

\newcommand{\eqref}[1]{(\ref{#1})}
\newcommand{\R}{\boldsymbol{\mathrm{R}}}
\newcommand{\A}{\EuScript{A}}
\newcommand{\gA}{\mathfrak{g}_{\EuScript{A}}}
\newcommand{\anchor}{\Psi}

\newcounter{example_counter}
\newcommand{\exmpl}{\addtocounter{example_counter}{1}{\sc EXAMPLE \arabic{example_counter}}.\,\,\,}
\newcounter{definition_counter}
\newcommand{\deff}{\addtocounter{definition_counter}{1}{\sc DEFINITION \arabic{definition_counter}}.\,\,\,}

\newcounter{theorem_counter}
\newcounter{remark_counter}
\newcommand{\rmrk}{\addtocounter{remark_counter}{1}{\sc REMARK \arabic{remark_counter}}.\,\,\,}

\begin{document}

\begin{frontmatter}

\title{Integrable partial differential equations and Lie--Rinehart algebras}

\author{Oleg I. Morozov}
\ead{oimorozov@gmail.com}
\address{Trapeznikov Institute of Control Sciences,
  \\
65 Profsoyuznaya Street,   Moscow 117997, Russia}

\begin{abstract}
We develop the method for  constructing Lax representations of {\sc pde}s via the twisted extensions of their algebras of contact symmetries by generalizing the construction to the Lie--Rinehart algebras. We  present examples of  application of the proposed technique.
\end{abstract}


\end{frontmatter}



\section{Introduction}

Theory of integrable partial differential equations  is an important part of modern ma\-the\-ma\-tics, and
numerous applications thereof are of big significance in physics. Lax re\-pre\-sen\-ta\-ti\-ons are widely
recognized as the key feature of integrable {\sc pde}s, be\-ing the starting point for such techniques as the
inverse scattering transformations, the bi-Ha\-mil\-to\-ni\-an structures, the B{\"{a}}cklund transformations, the recursion
operators, the nonlocal sym\-met\-ri\-es, the Darboux transformations, etc., see
\cite{WE,Zakharov82,RogersShadwick1982,NovikovManakovPitaevskiyZakharov1984,Konopelchenko1987,%
AblowitzClarkson1991,MatveevSalle1991,Olver1993,FokasGelfand1993,BacklundDarboux2001}
and references therein. Therefore the problem of finding intrinsic properties that ensure existence of a Lax
representation for a given {\sc pde} is of great interest. In the series of papers \cite{Morozov2017} ---
\cite{Morozov2021c} we proposed the method to attack this problem via the technique of the twisted extensions of
the Lie algebras of symmetries of  the {\sc pde}s under the study. This approach is of a limited scope and can not be
used in some examples. Analysis of  such examples reveals that the invariants of the symmetry algebras of both
the {\sc pde} {\it and} the Lax representation have to be included into the construction. This can be achieved
by considering the Lie--Rinehart algebras associated to the symmetry algebras of {\sc pde}s.

In the present  paper we generalize the approach of \cite{Morozov2017} --- \cite{Morozov2021c} for the
Lie--Rinehart algebras. We discuss the twisted extensions of the Lie--Rinehart algebras as well as the extensions by
appending an integral of a non-trivial 1-cocycle. Then we expose examples of con\-struc\-ting Lax representations
via these extensions of the Lie--Rinehart  algebras.

\section{Preliminaries and notation}

The presentation in this section closely follows
\cite{KrasilshchikVerbovetsky2011}---\cite{KrasilshchikVinogradov1989} and  \cite{VK1999}.
Let $\pi \colon \mathbb{R}^n \times \mathbb{R}^m \rightarrow \mathbb{R}^n$,
$\pi \colon (x^1, \dots, x^n, u^1, \dots, u^m) \mapsto (x^1, \dots, x^n)$, be a trivial bundle, and
$J^\infty(\pi)$ be the bundle of its jets of the infinite order. The local coordinates on $J^\infty(\pi)$ are
$(x^i,u^\alpha,u^\alpha_I)$, where $I=(i_1, \dots, i_n)$ are multi-indices with $i_k \ge 0$, and for every local section
$f \colon \mathbb{R}^n \rightarrow \mathbb{R}^n \times \mathbb{R}^m$ of $\pi$ the corresponding infinite jet
$j_\infty(f)$ is a section $j_\infty(f) \colon \mathbb{R}^n \rightarrow J^\infty(\pi)$ such that
$u^\alpha_I(j_\infty(f))
=\displaystyle{\frac{\partial ^{\#I} f^\alpha}{\partial x^I}}
=\displaystyle{\frac{\partial ^{i_1+\dots+i_n} f^\alpha}{(\partial x^1)^{i_1}\dots (\partial x^n)^{i_n}}}$.
We put $u^\alpha = u^\alpha_{(0,\dots,0)}$. Also, we will simplify notation in the following way: e.g., in the
case of $n=3$, $m=1$ we denote $x^1 = t$, $x^2= x$  $x^3= y$, 
and $u^1_{(i,j,k)}=u_{{t \dots t}{x \dots x}{y \dots y}}$ with $i$  times $t$, $j$  times $x$, and $k$ times $y$.

The  vector fields
\[
D_{x^k} = \frac{\partial}{\partial x^k} + \sum \limits_{\# I \ge 0} \sum \limits_{\alpha = 1}^m
u^\alpha_{I+1_{k}}\,\frac{\partial}{\partial u^\alpha_I},
\qquad k \in \{1,\dots,n\},
\]
$(i_1,\dots, i_k,\dots, i_n)+1_k = (i_1,\dots, i_k+1,\dots, i_n)$,  are called {\it total derivatives}.
They com\-mu\-te everywhere on
$J^\infty(\pi)$:  $[D_{x^i}, D_{x^j}] = 0$.

The {\it evolutionary vector field} associated to an arbitrary vector-valued smooth function
$\varphi \colon J^\infty(\pi) \rightarrow \mathbb{R}^m $ is the vector field
\[
\mathbf{E}_{\varphi} = \sum \limits_{\# I \ge 0} \sum \limits_{\alpha = 1}^m
D_I(\varphi^\alpha)\,\frac{\partial}{\partial u^\alpha_I}
\]
with $D_I=D_{(i_1,\dots\,i_n)} =D^{i_1}_{x^1} \circ \dots \circ D^{i_n}_{x^n}$.

A system of {\sc pde}s $F_r(x^i,u^\alpha_I) = 0$ of the order $s \ge 1$ with $\# I \le s$, $r \in \{1,\dots, R\}$ for some $R \ge 1$,
defines the submanifold $\EuScript{E}=\{(x^i,u^\alpha_I)\in J^\infty(\pi)\,\,\vert\,\,D_K(F_r(x^i,u^\alpha_I))=0,\,\,\# K\ge 0\}$
in $J^\infty(\pi)$.

A function $\varphi \colon J^\infty(\pi) \rightarrow \mathbb{R}^m$ is called a {\it (generator of an
infinitesimal) symmetry} of equation $\EuScript{E}$ when $\mathbf{E}_{\varphi}(F) = 0$ on $\EuScript{E}$. The
symmetry $\varphi$ is a solution to the {\it defining system}
\begin{equation}
\ell_{\EuScript{E}}(\varphi) = 0,
\label{defining_eqns}
\end{equation}
where $\ell_{\EuScript{E}} = \ell_F \vert_{\EuScript{E}}$ with the matrix differential operator
\[
\ell_F = \left(\sum \limits_{\# I \ge 0}\frac{\partial F_r}{\partial u^\alpha_I}\,D_I\right).
\]
The {\it symmetry algebra} $\mathrm{Sym} (\EuScript{E})$ of equation $\EuScript{E}$ is the linear space of
solutions to  (\ref{defining_eqns}) endowed with the structure of a Lie algebra over $\mathbb{R}$ by the
{\it Jacobi bracket} $\{\varphi,\psi\} = \mathbf{E}_{\varphi}(\psi) - \mathbf{E}_{\psi}(\varphi)$.
The {\it algebra of contact symmetries} $\mathrm{Sym}_0 (\EuScript{E})$ is the Lie subalgebra of $\mathrm{Sym} (\EuScript{E})$
defined as $\mathrm{Sym} (\EuScript{E}) \cap C^{\infty}(J^1(\pi))$. 

Let the linear space $\EuScript{W}$ be either $\mathbb{R}^N$ for some $N \ge 1$ or  $\mathbb{R}^\infty$
endowed with  local co\-or\-di\-na\-tes $w^a$, $a \in \{1, \dots , N\}$ or  $a \in  \mathbb{N}$, respectively.
Variables $w^a$ are cal\-led {\it pseudopotentials} \cite{WE}.  Locally, a {\it differential covering} of $\EuScript{E}$ is 
a trivial bundle $\varpi \colon J^\infty(\pi) \times \EuScript{W} \rightarrow J^\infty(\pi)$ equipped with {\it extended total derivatives}
\[
\widetilde{D}_{x^k} = D_{x^k} + \sum \limits_{a}
T^a_k(x^i,u^\alpha_I,w^b)\,\frac{\partial }{\partial w^a}
\]
such that $[\widetilde{D}_{x^i}, \widetilde{D}_{x^j}]=0$ for all $i \not = j$ if and only if $(x^i,u^\alpha_I) \in \EuScript{E}$. 
Define the partial derivatives of $w^a$ by  $w^s_{x^k} =  \widetilde{D}_{x^k}(w^s)$.  This yields the over-determined system
of {\sc pde}s 
\begin{equation}
w^a_{x^k} = T^a_k(x^i,u^\alpha_I,w^b)
\label{WE_prolongation_eqns}
\end{equation}
which is compatible if and only if  $(x^i,u^\alpha_I) \in \EuScript{E}$.
System \eqref{WE_prolongation_eqns} is referred to as the {\it covering equations}
or the {\it Lax representation} of equation $\EuScript{E}$.

Dually, the differential covering is defined by the
{\it Wahlquist--Estabrook forms}
\begin{equation}
\tau^a =d w^a - \sum \limits_{k=1}^{m} T^a_k(x^i,u^\alpha_I,w^b)\,dx^k
\label{WEfs}
\end{equation}
as follows: when $w^a$  and $u^\alpha$ are considered to be functions of $x^1$, ... , $x^n$, forms \eqref{WEfs}
are equal to zero if and only if  system \eqref{WE_prolongation_eqns} holds.

\section{Lie--Rinehart algebras and their extensions}

While {\'E}lie Cartan was well aware of the constructions underlying Lie--Rinehart algebras, see
\cite{Cartan1904}, at first time these algebras were introduced explicitly by J.-C. Herz \cite{Herz1953} under the
name of  `Lie pseudo-algebras'. Then they were examined by R. Palais \cite{Palais1961}   under the name
`d-Lie rings' and studied by G. Rinehart \cite{Rinehart1963}. The geometric counter-part of the  Lie--Rinehart
algebras are the Lie algebroids, see survey \cite{Mackenzie1995}.

The notion of the twisted Lie algebroid cohomology  was defined in \cite{Coueraud2017}. The first principle study of the LR algebra extensions were done (albeit, in a different language) in \cite{Huebschmann1999}. The extensive and proper study of the Lie algebroid/Lie--Rinehart algebra extensions were done in \cite{Bruzzo2015}  and (in full generality) in \cite{Aldrovandi2018}. The very natural LR algebra construction was proposed in the framework of the geometric approach to {\sc pde}s. These Lie algebroid/LRA structures (under the name "Der-modules") were introduced by A.M. Vinogradov, I.S. Krasil${}^{\prime}$shchik and V.V. Lychagin in their various  works in 1970--1986, see \cite{VKL1986} and references therein. This algebras naturally appear in geometry of jet spaces. The cohomology of $\mathrm{Der}$-complexes (including the extensions) were studied in 1980 thesis of V.N. Rubtsov and summarized in \cite{Rubtsov1980}.

In this section we follow \cite{Rinehart1963,Huebschmann1990,Mackenzie1995} in exposition of the basic definitions
of the theory of Lie--Rinehart algebras. Then we discuss the twisted extensions of these algebras as well as the extensions
by appending an integral of a non-trivial 1-cocycle.

\vskip 5 pt
\noindent
\deff
Let
$\R$ be a commutative ring, $\A$ be a commutative $\R$-algebra, and let $\gA$ be a Lie algebra
over $\R$ equipped with  two structures:
\begin{enumerate}
\item
a structure of a left $\EuScript{A}$-module on $\gA$, that is, a
map\footnote[7]{the unadorned tensor product symbol $\otimes$ will refer to the tensor product over $\R$.}
$\A \otimes \gA \rightarrow \gA$,
$a\otimes x \mapsto a \centerdot x$, such that
\[
(a \cdot b) \centerdot x = a \centerdot (b \centerdot x);
\]
\item
a map $\anchor \colon \gA \rightarrow \mathrm{Der} (\A)$ called the {\it anchor} which is a homomorphism of
Lie algebras over $\R$ and a homomorphism of $\A$-modules, that is
\begin{equation}
\anchor ([x,y]) = [\anchor(x),\anchor(y)]
\label{anchor_of_commutator}
\end{equation}
and
\[
\anchor(a\centerdot x)(b) = a \cdot (\anchor(x)(b))
\]
for $x, y \in \gA$ and $a, b \in \A$.
\end{enumerate}
Then $\gA$ is referred to as a {\it Lie--Rinehart algebra over} $\A$ provided there holds
\[
[x, a\centerdot x] =a \centerdot [x, y]  + \anchor(x)(a)\centerdot y.
\]

\vskip 5 pt

\noindent
\deff
A {\it Lie-Rinehart module over a Lie--Rinehart algebra} $\gA$ is a vector space $V$ equipped with
two operations
\[
\gA \otimes V \rightarrow V, \qquad x\otimes v \mapsto x(v)
\]
and
\[
\A \otimes V \rightarrow V, \qquad a\otimes v \mapsto a\centerdot v
\]
such that the first map makes $V$ into a Lie algebra module over the Lie $\R$-algebra $\gA$, while
the second map makes $V$ into an $\A$-module and additionally there hold
\[
(a\centerdot x) (v) = a\centerdot (x(v)),
\]
\[
x(a \centerdot v) = a \centerdot x(v) + \anchor(x)(a)\centerdot v.
\]

\vskip 5 pt

\noindent
\deff
Let $V$ be a Lie--Rinehart module over the Lie--Rinehart algebra $\gA$. Put $C^0(\gA, V) = V$ and
$C^k(\gA, V) = \mathrm{Hom}_{\A}(\Lambda^k(\gA), V)$ for $k \ge 1$. For $k \ge 0$ define differential 
\[
d\colon
C^k(\gA, V) \rightarrow C^{k+1}(\gA, V)
\]
by
\[
d \theta(x_1, ..., x_{k+1}) =
\sum\limits_{j=1}^{k+1}
(-1)^{j+1} \anchor (x_j)\,(\theta (x_1, ... ,\hat{x}_j, ... ,  x_{k+1}))
\]
\begin{equation}
\quad
+\sum\limits_{1\le i < j \le k+1} (-1)^{i+j+1}
\theta ([x_i,x_j],x_1, ... ,\hat{x}_p, ... ,\hat{x}_q, ... ,  x_{k+1}).
\label{Chevalley_Eilenberg_differential}
\end{equation}
The cohomology groups of the complex
\[
\fl
C^0(\gA, V) \stackrel{d}{\longrightarrow} C^1(\gA, V)
\stackrel{d}{\longrightarrow} \dots \stackrel{d}{\longrightarrow}
C^k(\gA, V) \stackrel{d}{\longrightarrow} C^{k+1}(\gA, V)
\stackrel{d}{\longrightarrow} \dots
\]
are
\[
H^k(\gA, V)
=
\frac{Z^k(\gA, V)}{B^k(\gA, V)}
=
\frac{\mathrm{ker}\,\, d \colon C^k(\gA, V) \rightarrow C^{k+1}(\gA, V)
}{\mathrm{im}\,\, d \colon C^{k-1}(\gA, V) \rightarrow C^{k}(\gA, V)}.
\]

\vskip 5 pt
\noindent
\rmrk
It is natural to consider $H^k(\gA, \A)$ as the cohomology groups of $\gA$ with trivial coefficients. These
groups will be  denoted as $H^k(\gA)$. Likewise, we denote $C^k(\gA, \A)=C^k(\gA)$, $Z^k(\gA, \A) =Z^k(\gA)$,
and $B^k(\gA, \A)=B^k(\gA)$.

\hfill $\diamond$

\vskip 5 pt

Below we consider Lie--Rinehart algebras within the following specific setting:
\begin{enumerate}
\item
$\R =\mathbb{R}$,
\item
$\A$  is the algebra of smooth or real-analytic functions $f(\pmb{w}) = f(w^1, \dots , w^n)$
defined on an open set $\EuScript{W} \subseteq \mathbb{R}^n$,
\item
the Lie algebra $\gA$ is a free $\A$-module with finite or countable set of generators $v_m$,    where
$m \in \{1, \dots  , M\}$ for some $M\ge 1$ or $m \in \mathbb{N}$. In the last case elements of $\gA$ are
linear combinations $\sum_m f_m(\pmb{w}) \,v_m$ with finite number of non-zero functions $f_m$.
\end{enumerate}
Commutators of the basis elements
\begin{equation}
[v_i, v_j] = \sum \limits_{k} c^k_{ij}(\pmb{w})\,v_k
\label{commutator_v_i_v_j}
\end{equation}
define the {\it structure functions}  $c^k_{ij}(\pmb{w})$, and the anchor
has the form
\begin{equation}
\anchor (v_i) = \sum\limits_{q=1}^{n} h_i^q(\pmb{w})\,\partial_{w^q}
\label{anchor_v_i}
\end{equation}
for some functions $h^q_i(\pmb{w})$. The skew-symmetry of commutator entails
$c^k_{ij}(\pmb{w})= -c^k_{ji}(\pmb{w})$.
The Jacobi identity $\sum_{\mathrm{cycl}(i,j,k)} [v_i, [v_j, v_k]]=0$ gives
\[
\sum_{\mathrm{cycl}(i,j,k)} \left(\sum \limits_{q} h^q_i \,\partial_{w^q} c^m_{jk}
+ \sum\limits_{l} c^{l}_{jk}\,c^m_{li}\right) =0,
\]
while from \eqref{anchor_of_commutator} it follows that
\[
\sum \limits_{s} \left(h^s_i\,\partial_{w^s} h^q_j - h^s_j\,\partial_{w^s} h^q_i\right) =
\sum \limits_{k}
c^k_{ij} \,h^q_k.
\]

Consider $\A$-linear functions  $\theta^i \colon \gA \rightarrow \A$ defined by  $\theta^i(v_j) = \delta^i_j$.
Then  \eqref{Chevalley_Eilenberg_differential}, \eqref{commutator_v_i_v_j}, and \eqref{anchor_v_i}
 yield the {\it structure equations}
\[
\left\{
\begin{array}{lcl}
d\theta^i &=& \displaystyle{-\sum \limits_{j<k} c^i_{jk}(\pmb{w}) \,\theta^j \wedge \theta^k},
\\
dw^q &=& \displaystyle{\sum \limits_{i} h^q_i(\pmb{w})\,\theta^i}.
\end{array}
\right.
\]
{\it of the Lie--Rinehart algebra} $\gA$.

\vskip 4 pt

In all the examples below the image of the anchor is finite-dimensional, in other words, the sums in the
{\sc rhs} of equations for $dw^q$ are finite. For such Lie--Rinehart algebras  we can assume without loss of
generality that $\mathrm{rank} \left(h^q_i\right) = n = \mathrm{dim} \EuScript{W}$, since otherwise we can
reduce the number of functionally independent variables $w^q$. We rename $\sum_i h^q_i\,\theta^i =\colon \eta^q$,
then we have $dw^q = \eta^q$ and $d\eta^q=0$, so $B^1(\gA) = \langle \eta^1, \dots , \eta^n\rangle$.
Furthermore,  for a Lie--Rinehart algebra with the finite-dimensional image of the anchor we can write the
structure equations in the form
\[
\left\{
\begin{array}{lcl}
d\vartheta^i &=&
\displaystyle{
\sum \limits_{j<k} P^i_{jk}(\pmb{w}) \,\vartheta^j \wedge \vartheta^k
+ \sum \limits_{j,q} Q^i_{jq}(\pmb{w}) \,\vartheta^j \wedge \eta^q
+\sum \limits_{q <s} R^i_{qs}(\pmb{w}) \,\eta^q \wedge \eta^s,}
\\
d\eta^q &=& 0,
\\
dw^q &=&\eta^q
\end{array}
\right.
\]
with some functions $P^i_{jk}$,  $Q^i_{jq}$, $R^i_{qs}$ and 1-forms $\vartheta^i$ such that collection
$\{\eta^q, \vartheta^i\}$ provides a basis for $C^1(\gA)$.

\vskip 5 pt
\noindent
\deff
Consider a Lie--Rinehart  algebra $\gA$ with $H^1(\gA) \neq \{[0]\}$. Let $\alpha$ be a non-trivial 1-cocycle,
that is, $d\alpha = 0$ and $\alpha \not \in B^1(\gA)$. For a constant $c \in \mathbb{R}$ define the
{\it twisted differential} $d_{c\alpha}\colon C^k(\gA) \rightarrow C^{k+1}(\gA)$ by the formula
\[
d_{c\alpha} \theta = d\theta -c\,\alpha \wedge \theta.
\]
Then $d^2_{c\alpha}=0$. The cohomology groups $H^{*}_{c\alpha} (\gA)$ of the associated complex are re\-fer\-red
to as the {\it twisted cohomology groups} of $\gA$.

\vskip 5 pt
\noindent
\deff
Suppose $H^2_{c\alpha} (\gA) \neq \{[0]\}$ for some $c \in \mathbb{R}$ and $\Omega$ is a non-trivial twisted
2-cocycle.Then equation
\begin{equation}
d\sigma = c\,\alpha \wedge \sigma +\Omega
\label{sigma_def_equation}
\end{equation}
with unspecified 1-form $\sigma$ is compatible with the structure equations of $\gA$. The Lie--Rinehart algebra
$\tilde{\mathfrak{g}}_{\A}$ with the structure equations obtained by appending \eqref{sigma_def_equation} to the
structure equations of $\gA$ is referred to as the {\it twisted extension} of $\gA$.

\vskip 5 pt
\noindent
\exmpl
Consider the Lie--Rinehart algebra
\[
\gA = \left\{\,\, \sum \limits_{k=1}^{4} f_k(w)\,v_k \,\,\vert \,\, f_k \in C^{\infty}(\mathbb{R})\,\,\right\}
\]
over $\A = C^{\infty}(\mathbb{R})$ with non-zero commutators
\[
[v_1, v_2] =-v_2, \qquad [v_1, v_3] = - v_3, \qquad [v_2, v_4] = -v_3
\]
of the basis elements $v_1$, ... , $v_4$ and the anchor
\[
\anchor(v_k) =
\left\{
\begin{array}{lcl}
0, &~& 1 \le k \le 3,
\\
\partial_w, && k=4.
\end{array}
\right.
\]
The structure equations of $\gA$ read
\begin{equation}
\left\{
\begin{array}{lcl}
d\theta^1 &=& 0,
\\
d\theta^2 &=& \theta^1 \wedge \theta^2,
\\
d\theta^3 &=& \theta^1 \wedge \theta^3+\theta^2 \wedge \theta^4,
\\
d\theta^4 &=& 0,
\\
dw &=& \theta^4.
\end{array}
\right.
\label{toy_algebra_se}
\end{equation}
We have $H^1(\gA) = \{[\theta^1]\}$, and  the straightforward computations give
\[
H^2_{c\theta^1}(\gA) =
\left\{
\begin{array}{lcl}
\langle [\theta^1 \wedge \theta^2], [\theta^1 \wedge (w\,\theta^2+\theta^3)] \rangle, &~& c=1,
\\
\langle [\theta^2 \wedge \theta^3]\rangle, &~& c=2,
\\
0, && c \not\in \{1,2\}.
\end{array}
\right.
\]
Therefore we have the three-dimensional twisted extension of $\gA$  defined by appending  equations
\[
\left\{
\begin{array}{lcl}
d\sigma^1 &=& \theta^1 \wedge \sigma^1 + \theta^1 \wedge \theta^2,
\\
d\sigma^2 &=& \theta^1 \wedge \sigma^2 + \theta^1 \wedge (w\,\theta^2+\theta^3),
\\
d\sigma^3 &=& 2\,\theta^1 \wedge \sigma^3+\theta^2 \wedge \theta^3
\end{array}
\right.
\]
to system \eqref{toy_algebra_se}. Then in the basis $\langle v_1, \dots  , v_7 \rangle$
dual to forms $\theta^k$, $\sigma^j$ the non-zero com\-mu\-ta\-tors for the extended Lie--Rinehart algebra are
\[
[v_1,v_2] =-v_2-v_5-w\,v_6,
\quad
[v_1,v_3]=-v_3-v_6,
\quad
[v_1,v_5]=-v_5,
\]
\[ [v_1,v_6]=-v_6,
\quad
[v_1,v_7]=-2\,v_7,
\quad
[v_2,v_3]=-v_7,
\quad
[v_2,v_4]=-v_3,
\]
and for the anchor we have $\anchor(v_k) =0$ when $k \in \{5, 6, 7\}$.

\hfill $\diamond$

\vskip 5 pt
\noindent
\deff
Suppose we have $H^1(\gA) \neq \{[0]\}$ for a Lie--Rinehart algebra $\gA$, and  $\alpha$ is a non-trivial
1-cocycle on $\gA$. Then we extend $\A$ and thus $\gA$ by considering algebra
$\tilde{\A} = C^{\infty}(\EuScript{W} \times\mathbb{R})$ of functions $f(w^1, \dots, w^{n+1})$ and extending
the anchor by $dw^{n+1} =\alpha$. We refer this extension as {\it appending an integral of} $\alpha$.
Notice that $\alpha \in B^1(\mathfrak{g}_{\tilde{\A}})$.

\vskip 5 pt
\noindent
\rmrk
The procedure of extension by appending an integral of a 1-cocycle is applicable to a Lie algebra over
$\mathbb{R}$  with non-trivial first cohomology group. If $H^1(\mathfrak{a}) \neq {0}$ for a  Lie algebra
$\mathfrak{a}$ and  $\alpha$ is a non-trivial 1-cocycle, then  the extended algebra is the Lie--Rinehart
algebra  $\mathfrak{a}_{C^{\infty}(\mathbb{R})}$, where $C^{\infty}(\mathbb{R})$ consists of smooth functions
$f(w)$ of $w \in \mathbb{R}$ and the structure equations of $\mathfrak{a}_{C^{\infty}(\mathbb{R})}$ are obtained
by appending equation $dw =\alpha$ to the structure equations of $\mathfrak{a}$.

\vskip 5 pt
\noindent
\deff
For a Lie--Rinehart algebra $\gA$ with  non-trivial  second twisted co\-ho\-mo\-lo\-gy group  we can combine the
procedures of twisted extension and appending an integral. Namely, if $\alpha$ is a non-trivial 1-cocycle and
$\Omega$ is non-trivial twisted 2-cocycle with $d\Omega=c\,\alpha\wedge \Omega$ for $c\in\mathbb{R}$, we define
the combined extension  of $\gA$ in two steps: first, constructing the twisted extension
$\tilde{\mathfrak{g}}_{\A}$ of $\gA$, and then extending $\A$ to $\tilde{\A}$ by appending an integral $w$ of
1-cocycle $\alpha$. The resulting Lie--Rinehart algebra $\tilde{\mathfrak{g}}_{\tilde{\A}}$ is not a twisted
extension of $\gA$ anymore, since $\alpha \in B^1(\tilde{\mathfrak{g}}_{\tilde{\A}})$. The structure equations of
$\tilde{\mathfrak{g}}_{\tilde{\A}}$ are obtained from the structure equations of $\gA$ by adding equations
$d\sigma =c\,\alpha \wedge \sigma +\Omega$ and $dw=\alpha$.

\section{Lax representations via extensions of Lie--Rinehart algebras}

In this section we expose three examples of constructing Lax representations via the pro\-ce\-du\-res of the
combined extension of a Lie--Rinehart algbera and extension of a Lie algebra by appending an integral of a
non-trivial 1-cocycle. To the best of our knowledge the results of Examples 2 and 3 can not be recovered by the method of
\cite{Morozov2017}.  Example 4 exposes new Lax re\-pre\-sen\-ta\-ti\-on for the  hyper-CR equation of Ein\-stein--Weyl struc\-tu\-res \eqref{MPD}.

\vskip 7 pt
\noindent
\exmpl
Consider equation $\EuScript{E}_1$
\begin{equation}
u_{yy} = \frac{u_{tx}}{u_{xy}}+F(u_{xx})\,u_{xy}^2,
\label{Pavlov_Chazy_eq}
\end{equation}
where function $F$ is a solution to Chazy's equation
\begin{equation}
F^{\prime\prime\prime}  +12\,F\,F^{\prime\prime}-18\,(F^{\prime})^2 = 0.
\label{Chazy_ODE}
\end{equation}
Equation \eqref{Pavlov_Chazy_eq} was introduced in \cite{Pavlov2003}, the Lax representation thereof was
presented in \cite{Pavlov2004} in implicit form and in \cite{CleryFerapontov2020} in explicit form.

The algebra $\mathrm{Sym}_0(\EuScript{E}_1)$ of contact symmetries for equation \eqref{Pavlov_Chazy_eq} has
generators\footnote[7]{We carried out computations of generators of contact symmetries in the {\it Jets}
software \cite{Jets}.}
\[
\varphi_0(A_0) = -A_0\,u_t-\frac{1}{3}\,A_0^{\prime}\,y\,u_y-\frac{1}{18}\,A_0^{\prime\prime}\,y^3,
\]
\[
\varphi_1(A_1)=-A_1\,u_y -\frac{1}{2}\,A_1^{\prime}\,y^2,
\]
\[
\varphi_2(A_2) = A_2\,y,
\]
\[
\varphi_3(A_3)=A_3,
\]
\[
\psi_0 = 3\,u-\frac{3}{2}\,x\,u_x-y\,u_y,
\]
\[
\psi_1 = -u_x,
\]
\[
\psi_2 = x.
\]
where $A_i=A_i(t)$ are arbitrary smooth functions of $t$.  The action of  $\mathrm{Sym}_0(\EuScript{E}_1)$ on
$J^2(\pi)$ with $\pi \colon (t,x,y,u) \mapsto (t,x,y)$  has two invariants $u_{xx}$ and
$(u_{xy}\,u_{yy}-u_{tx})\,u_{xy}^{-3}$. These invariants are functionally dependent when restricted to
$\EuScript{E}_1$: $(u_{xy}\,u_{yy}-u_{tx})\,u_{xy}^{-3} = F(u_{xx})$. Using the technique of moving frames
\cite{Olver_Pohjanpelto_2005,Cheh_Olver_Pohjanpelto_2005,Olver_Pohjanpelto_Valiquette_2009}
the structure equations of $\mathrm{Sym}_0(\EuScript{E}_1)$ can be written in the form
\begin{equation}
\left\{
\begin{array}{lcl}
d\alpha_0 &=&0,
\\
d\alpha_1 &=&\alpha_0 \wedge \alpha_1,
\\
d\alpha_2 &=& \alpha_0 \wedge \alpha_2 - \eta \wedge \alpha_1,
\\
d\eta &=& 0,
\\
d\Theta &=&
h_0\,\alpha_0 \wedge \partial_{h_0}\,\Theta+
\partial_{h_1} \Theta \wedge \left(\Theta - \frac{2}{3}\,h_0\,\partial_{h_0}\,\Theta\right),
\\
d\theta_{3,-1} &=&
2\,\alpha_0 \wedge \theta_{3,-1}+ \theta_{3,0} \wedge \theta_{0,0}
+\frac{1}{3}\,\theta_{2,0} \wedge \theta_{1,0} +\alpha_1 \wedge \alpha_2,
\\
dw &=& \eta,
\end{array}
\right.
\label{se_of_Pavlov_Chazy_eq}
\end{equation}
where
\[
\Theta = \sum \limits_{k=0}^{3} \sum \limits_{m=0}^{\infty} \frac{1}{m!} \,h_0^k\,h_1^m\,\theta_{k,m},
\]
$h_0^k =0$ when $k >3$, $dh_i=0$, and $w= u_{xx}$.
Equations for $d\alpha_0$, $d\alpha_2$, $d\alpha_2$, $d\eta$, and $dw$ differ only in notation from system
\eqref{toy_algebra_se}, therefore,  according to Example 1 and Definition 7, the Lie--Rinehart algebra with the
structure equations \eqref{se_of_Pavlov_Chazy_eq} admits the combined extension whose structure equations are
obtained by appending equations
\begin{equation}
d\sigma = \alpha_0 \wedge \sigma +\alpha_0 \wedge \alpha_1
\label{dtau}
\end{equation}
and
\[
dq = \alpha_0
\]
to system \eqref{se_of_Pavlov_Chazy_eq}. In these equations $\sigma$ is an unspecified 1-form and $q$ is new
invariant. In what follows we need explicit expressions for the Maurer--Cartan forms
\[
\alpha_1 = \mathrm{e}^q\,dx,
\]
\[
\alpha_2 = \mathrm{e}^q\,(du_x-u_{xx}\,dt),
\]
\[
\eta = du_{xx},
\]
\[
\theta_{0,0} = \mathrm{e}^q\,u_{xy}^3\,dt,
\]
\[
\theta_{1,0} = \mathrm{e}^q\,\left(u_{xy}\,dy +(u_{tx}-2\,F\,u_{xy}^3)\,dt\right),
\]
\[
\theta_{3,-1} = \mathrm{e}^{2q}\,(du-u_t\,dt-u_x\,dx - u_y\,dy).
\]
Integration of equation \eqref{dtau} yields
\[
\sigma = \mathrm{e}^{q}\,(dv +q\,dx).
\]
To find the Wahlquist--Estabrook form of a Lax representation for equation \eqref{Pavlov_Chazy_eq}  we
con\-si\-der the  linear combination
\[
\fl
\sigma-P_1\,\theta_{0,0} -P_2\,\theta_{1,0} =
\mathrm{e}^q\,\left(
dv +q\,dx-P_2\,u_{xy}\,dy - (P_1\,u_{xy}^3+P_2\,(u_{tx}-2\,F\,u_{xy}^3)\,dt\right),
\]
where  coefficients $P_i$ are functions of invariants $u_{xx}$ and $q$. This 1-form defines the Lax
representation
\begin{equation}
\left\{
\begin{array}{lcl}
v_t &=& P_1\,u_{xy}^3 +P_2\,(u_{tx}-2\,F\,u_{xy}^3),
\\
v_y &=& P_2\,u_{xy}
\end{array}
\right.
\label{covering_of_Pavlov_Chazy_eq}
\end{equation}
provided
$q =  -v_x$ and thus $P_i=P_i(u_{xx},v_x)$.
System \eqref{covering_of_Pavlov_Chazy_eq} differs only in notation from the Lax representation found in
\cite{CleryFerapontov2020}.
Analysis of compatibility of \eqref{covering_of_Pavlov_Chazy_eq} yields
\[
P_1 = \frac{1}{2}\,(P_{2,u_{xx}}+P_2\,P_{2,v_x})+2\,P_2\,F
\]
and the over-determined system
\[\fl
P_{2,v_xv_x} =
\frac{2\,P_{2,v_x}^3-F_{u_{xx}u_{xx}}-6\,F_{u_{xx}}\,P_{2,v_x}-6\,F\,P_{2,v_x}^2}
{P_{2,u_{xx}}+P_2\,P_{2,v_x}},
\]
\[\fl
P_{2,v_x,u_{xx}} =
\frac{
P_2\,F_{u_{xx}u_{xx}}
+3\,F_{u_{xx}}\,(P_2\,P_{2,v_x}-P_{2,u_{xx}})-6\,F\,P_{2,v_x}\,P_{2,u_{xx}}
+2\,P_{2,v_x}^2\,P_{2,u_{xx}}
}
{P_{2,u_{xx}}+P_2\,P_{2,v_x}},
\]
\[\fl
P_{2,u_{xx}u_{xx}} = \frac{2\,P_{2,v_x}\,P_{2,u_{xx}}^2-6\,F\,P_{2,u_{xx}}^2
-P_2^2\,F_{u_{xx}u_{xx}}
+6\,P_2\,F_{u_{xx}}\,P_{2,u_{xx}}}{P_{2,u_{xx}}+P_2\,P_{2,v_x}}
\]
for function $P_2$. In its turn this system is compatible if and only if  equation \eqref{Chazy_ODE} holds.

\hfill $\diamond$

\vskip 5 pt
\noindent
\rmrk
While each equation $(u_{xy}\,u_{yy}-u_{tx})\,u_{xy}^{-3} = G(u_{xx})$ with an arbitrary function $G$ admits 
$\mathrm{Sym}_0(\EuScript{E}_1)$ as the symmetry algebra,  this equation possesses the Lax representation if and only if
$G$ is a solution to Chazy's {\sc ode} \eqref{Chazy_ODE}, cf. \cite{CleryFerapontov2020}.  

\hfill $\diamond$

\vskip 7 pt
\noindent
\exmpl
Equation $\EuScript{E}_2$
\begin{equation}
u_{yy} = u_y\,(u_{ty} +u_x\,u_{xy}-u_y\,u_{xx})
\label{E_3}
\end{equation}
was introduced in \cite{Morozov2012}. Algebra $\mathrm{Sym}_0(\EuScript{E}_2)$ of contact symmetries of this equa\-ti\-on 
 is ge\-ne\-ra\-ted by functions
\[
\varphi_0(A_0) = -A_0\,u_t-A_0^{\prime}\,x\,u_x+A_0^{\prime}\,u+\frac{1}{2}\,A_0^{\prime\prime}\,x^2,
\]
\[
\varphi_1(A_1) = -A_1\,u_x+A_1^{\prime}\,x,
\]
\[
\varphi_2(A_2) = A_2,
\]
\[
\psi_0 = -y\,u_y,
\]
\[
\psi_1 = -u_y,
\]
where $A_i=A_i(t)$ are arbitrary functions of $t$. The structure equations of $\mathrm{Sym}_0(\EuScript{E}_2)$
can be written in the form
\[
\left\{
\begin{array}{lcl}
d\alpha_0 &=& 0,
\\
d\alpha_1 &=& \alpha_0 \wedge \alpha_1,
\\
d\Theta &=& \partial_{h_1} \Theta \wedge \Theta,
\end{array}
\right.
\]
where
\[
\Theta = \sum \limits_{k=0}^{2} \sum \limits_{m=0}^{\infty} \frac{1}{m!} \,h_0^k\,h_1^m\,\theta_{k,m},
\]
$h_0^k =0$ for $k >2$, and $dh_i=0$.  From these equations it follows that
\[
H^1(\mathrm{Sym}(\EuScript{E}_2)) = \langle[\alpha_0]\rangle
\]
and
\[
H^2_{c\alpha_0}(\mathrm{Sym}(\EuScript{E}_2)) =
\left\{
\begin{array}{lcl}
\langle[\alpha_0 \wedge \alpha_1]\rangle, &~& c=1,
\\
\{[0]\}, && c\neq 1.
\end{array}
\right.
\]
The non-trivial twisted 2-cocycle defines the twisted extension of the Lie algebra $\mathrm{Sym}(\EuScript{E}_2)$
with the additional structure equation
\[
d\sigma = \alpha_0 \wedge \sigma+\alpha_0 \wedge \alpha_1.
\]
In accordance with Remark 2 the obtained Lie algebra admits extension by appending integral of $\alpha_0$.
The resulting  Lie--Rinehart algebra has the following Maurer--Cartan forms
\[
\alpha_0 = dq,
\]
\[
\alpha_1 = \mathrm{e}^q\,dy,
\]
\[
\theta_{0,0}= u_y^{-1}\,\mathrm{e}^q\,dt,
\]
\[
\theta_{1,0} = u_y^{-1}\,\mathrm{e}^q\,(dx-u_x\,dt),
\]
\[
\theta_{2,0} = u_y^{-1}\,\mathrm{e}^q\,(du-u_t\,dt-u_x\,dx),
\]
\[
\sigma = \mathrm{e}^q\,\left(dv+q\,dy\right).
\]
Consider the linear combination
\[
\tau=\sigma-Q_1\,\theta_{1,0} - Q_2\,\theta_{0,0} =
\mathrm{e}^{q}\,\left(
dv -\frac{Q_1}{u_y}\,dx -\frac{Q_2-Q_1\,u_x}{u_y}\,dt+q\,dy\right),
\]
where $Q_i$ are functions of $q$.
Upon setting $\tau = 0$ we obtain the over-determined system for function $v=v(t,x,y)$. This system yields
$q = -v_y$ and hence $Q_i=Q_i(v_y)$. Analysis of compatibility of two other equations
\begin{equation}
\left\{
\begin{array}{lcl}
v_t &=& \displaystyle{\frac{Q_2-Q_1\,u_x}{u_y}},
\\
v_x &=& \displaystyle{\frac{Q_1}{u_y}}.
\end{array}
\right.
\label{E_3_covering_a}
\end{equation}
of the system gives
\begin{equation}
Q_1= \frac{1}{\Phi^{\prime}},
\qquad Q_2 =\frac{\Phi}{\Phi^{\prime}},
\label{E_3_covering_b}
\end{equation}
where $\Phi=\Phi(v_y)$ is a solution to {\sc ode}
\begin{equation}
\Phi^{\prime\prime} = \Phi\,(\Phi^{\prime})^2.
\label{E_3_covering_c}
\end{equation}
Up to a change of notation this equation defines function $\Phi(v_y)$ implicitly by formula
\[
v_y = \mathrm{erf}(\Phi) =
\frac{2}{\sqrt{\pi}}\,\int \limits_{0}^{\Phi} \mathrm{e}^{-z^2}\,dz.
\]
In another notation system \eqref{E_3_covering_a}, \eqref{E_3_covering_b}, \eqref{E_3_covering_c} was found in \cite{Morozov2012} by the method of contact integrable extensions pro\-po\-sed in \cite{Morozov2009}.

\hfill $\diamond$

\vskip 5 pt
\noindent
\exmpl
Consider the hyper-CR equation of Ein\-stein--Weyl struc\-tu\-res $\EuScript{E}_3$ 
\begin{equation}
u_{yy}=u_{tx}+u_y\,u_{xx}-u_x\,u_{xy}.
\label{MPD}
\end{equation}
introduced independently in \cite{Mikhalev1992,Pavlov2003b,Dunajski2004}, where an `isospectral' Lax re\-pre\-sen\-ta\-ti\-on for this equation was found.  As we show in \cite{Morozov2019,Morozov2021a}, this Lax re\-pre\-sen\-ta\-ti\-on as well as its `nonisospectral' generalization  can be derived from  the  twisted extension of the symmetry algebra $\mathrm{Sym}(\EuScript{E}_3)$ of \eqref{MPD}. In this example we apply the technique described in Remark 2 to find the further generalization of the Lax representation from \cite{Morozov2021a}.   
  
As we show in \cite{Morozov2019}, the structure equations of  the Lie algebra $\mathrm{Sym}(\EuScript{E}_3)$
read
\[   
\left\{
\begin{array}{lcl}
d\alpha_0 &=& 0,
\\
d\alpha_1 &=& \alpha_0 \wedge \alpha_1,
\\
d \Theta &=& \nabla_1 (\Theta) \wedge \Theta + (h_0\,\alpha_0 + h_0^2\,\alpha_1) \wedge \nabla_0 (\Theta),
\end{array}
\right.
\] 
where
\[   
\Theta = \sum \limits_{k=0}^{3} \sum \limits_{m=0}^{\infty} \frac{h_0^kh_1^m}{m!}\,\theta_{k,m},
\] 
with the formal parameters $h_0$ and $h_1$ such that $h_0^k =0$ when $k>3$.  The additional structure equation for the twisted extension of $\mathrm{Sym}(\EuScript{E}_3)$ has the form
\[   
d \sigma = \alpha_0 \wedge \sigma + \alpha_0 \wedge \alpha_1.
\]  
Just as in papers \cite{Morozov2019,Morozov2021a}, we need the following  Maurer--Cartan forms for constructing the Lax representations of 
equation \eqref{MPD}: 
$\alpha_0 = dq$, $\alpha_1 = - \mathrm{e}^q\, ds$,
$\theta_{0,0} = r\,dt$,
$\theta_{1,0} = r\, \mathrm{e}^q\,(dy - (u_x-2\,s)\,dt)$,
$\theta_{2,0}= r\, \mathrm{e}^{2q}\,(dx -(u_x-s)\,dy - (u_y+s\,u_x-s^2)\,dt)$,
$\theta_{3,0} = r\, \mathrm{e}^{3q}\,(du - u_t\,dt - u_x\,dx -u_y\,dy)$,
and $\sigma  = \mathrm{e}^q\,(dv-q \,ds)$, 	where $q$, $s$, $v$, and $r$  are free parameters.
We choose the linear combination
\[
\tau =
\sigma - \sum \limits_{k=0}^2 S_k\,\theta_{k,0} =
\mathrm{e}^q\,\left(
dv - q\,ds - S_2\,r\,\mathrm{e}^q\,dx
-r\,(S_1+S_2\,\mathrm{e}^q (s-u_x))\,dy
\right.
\]
\[
\qquad\left.
-r\,(S_0\,\mathrm{e}^{-q}+S_1\,(2\,s-u_x)+S_2\,\mathrm{e}^q (s^2-s\,u_x-u_y))\,dt
\right)
\]
of the form $\sigma$ and the basic horizontal forms $\theta_{0,0}$, $\theta_{1,0}$, $\theta_{2,0}$
as the Wahlquist--Estabrook form of a Lax  representation.
Unlike the computations in [24,25], we now treat coefficients $S_k$ as functions of
 the integral $q$
of form $\alpha_0 \in H^1(\mathrm{Sym}(\EuScript{E}_3))$
rather than constants.
%
%
Since the restriction of form $\tau$ to the sections of the bundle $(t,x,y,u,v) \mapsto (t,x,y)$  has to be equal to zero, we put $q=v_s$.  By renaming $r$ we obtain without loss of generality $S_2=1$ and  $r= v_x\,\exp\, (-v_s)$.
Then the form
\[
\fl
\tau=
\mathrm{e}^q\,\left(
dv - v_s\,ds - v_x\,(dx+(s-u_x+S_1\,\mathrm{e}^{-v_s})\,dy
\right.
\]
\[
\qquad\left.
+(s^2-s\,u_x-u_y+S_1\,\mathrm{e}^{-v_s}\,(2\,s-u_x)+S_0\,\mathrm{e}^{-2v_s})\,dt)
\right)
\]
is equal to zero whenever there hold
\begin{equation}
\left\{
\begin{array}{lcl}
v_t &=& \left(s^2-s\,u_x-u_y+S_1\,\mathrm{e}^{-v_s}\,(2\,s-u_x)+S_0\,\mathrm{e}^{-2v_s}\right)\,v_x,
\\
v_y &=& \left(s-u_x+S_1\,\mathrm{e}^{-v_s}\right)\,v_x.
\end{array}
\right.
\label{generalized_covering}
\end{equation}
Just as in paper \cite{Morozov2021a},  the analysis of compatibility  condition $(v_t)_y=(v_y)_t$ 
for system \eqref{generalized_covering}  leads to $S_0=S_1^2$.
Denoting $R=S_1\,\mathrm{e}^{-v_s}$ we obtain the Lax representation
\begin{equation}
\left\{
\begin{array}{lcl}
v_t &=& \left(s^2-s\,u_x-u_y+R\,(2\,s-u_x)+R^2\right)\,v_x,
\\
v_y &=& \left(s-u_x+R\right)\,v_x
\end{array}
\right.
\label{new_covering}
\end{equation}
of equation (\ref{MPD})   with an arbitrary function $R=R(v_s)$. When $R=0$, this system coincides with the Lax representation from \cite{Mikhalev1992,Pavlov2003b,Dunajski2004}, while when $R=\mathrm{e}^{-v_s}$ we get the Lax representation from \cite{Morozov2021a}.   

\hfill $\diamond$

\section{Concluding remarks}
We have proposed the generalization of the method for constructing Lax representations  based on twisted extensions of Lie algebras to the Lie-Ri\-ne\-hart  algebras and showed that new technique allows one to recover in a simple manner known results as well as to find new Lax representations. We hope that further examples will clarify this technique and the limits of its  applicability.  The very important issue to address in the future research is to establish relationship between extensions of Lie-Rinehart algebras and the method of contact integrable extensions of Lie symmetry pseudo-groups pro\-po\-sed in \cite{Morozov2009}.

\section*{Acknowledgments}

I would like to thank  J.-H. Chang for his warm hospitality during my visit to Taipei,  where a part of the work  was performed.
I am very grateful to B.S. Kruglikov for sti\-mu\-la\-ting discussions. 
I would like to express my sincere gratitude  to I.S. Kra\-{}sil${}^{\prime}$\-{}shchik for  important comments.

I would like to thank the anonymous referees for several useful suggestions that helped improve the exposition of the paper.


\end{document}